\documentclass{elsarticle}

\usepackage{graphicx}

\usepackage{amsmath}
\usepackage{amssymb}
\usepackage{mathdots}

\biboptions{sort&compress}

\journal{Computer Physics Communications}

\def\G{{\bf G}}
\def\k{{\bf k}}
\def\q{{\bf q}}
\def\b{{\bf b}}
\def\R{{\bf R}}
\def\0{{\bf 0}}

\def\ket#1{\vert#1\rangle}
\def\exv#1#2#3{\langle#1\vert#2\vert#3\rangle}
\def\ee{\mathrm{e}}

\begin{document}

\begin{frontmatter}



\title{Automated quantum conductance calculations using
  maximally-localised Wannier functions} 

\author[Imperial]{Matthew Shelley}
\author[MIT]{Nicolas Poilvert}
\author[Imperial]{Arash A.~Mostofi\corref{cor1}}
\ead{a.mostofi@imperial.ac.uk}
\author[Oxford]{Nicola Marzari}

\cortext[cor1]{Corresponding author}

\address[Imperial]{The Thomas Young Centre for Theory and Simulation of
  Materials, Imperial College London, London SW7 2AZ, UK}
\address[MIT]{Department of Materials Science and Engineering,
  Massachusetts Institute of Technology, Cambridge MA 02139, USA}
\address[Oxford]{Department of Materials,
  University of Oxford, 16 Parks Road, Oxford OX1 3PH, UK}

\begin{abstract}
A robust, user-friendly, and automated method to determine quantum
conductance in quasi-one-dimensional systems is
presented. The scheme relies upon an initial density-functional
theory calculation in a specific geometry after which the ground-state
eigenfunctions are transformed to a maximally-localised Wannier
function (MLWF) basis. In this basis, our novel algorithms manipulate
and partition the Hamiltonian for the calculation of coherent electronic
transport properties within the Landauer-Buttiker formalism. Furthermore,
we describe how short-ranged 
Hamiltonians in the MLWF basis can be combined to build model
Hamiltonians of large ($>$10,000 atom) disordered systems without loss
of accuracy. These automated algorithms have been implemented in the
Wannier90 code~\cite{mostofi_2008}, which is interfaced to a number of
electronic structure codes such as Quantum-ESPRESSO, AbInit, Wien2k,
SIESTA and FLEUR. We apply our methods to an Al atomic chain with a Na
defect, an axially heterostructured Si/Ge nanowire and to
a spin-polarised defect on a zigzag graphene nanoribbon.
\end{abstract}


\end{frontmatter}



\section{Introduction}
\label{sec:intro}

Nanostructured materials, such as carbon nanotubes and silicon
nanowires, promise advances in wide-ranging device applications such
as photonics\cite{cui_2001}, thermoelectrics
\cite{hochbaum_2008, boukai_2008} and biological/chemical 
sensing\cite{atwater_2010}. Successful
incorporation of such structures in real devices requires bottom-up
approaches to design, which in turn, require an understanding
of electronic transport at the nano and mesoscales. 

First-principles simulations based on density-functional
theory (DFT) are now well-established as a powerful tool for materials 
modelling. Their success is largely due to the high
accuracy and computational efficiency that can be obtained for
many classes of materials. 

\begin{figure}
\centering
\includegraphics[height=2.8cm]{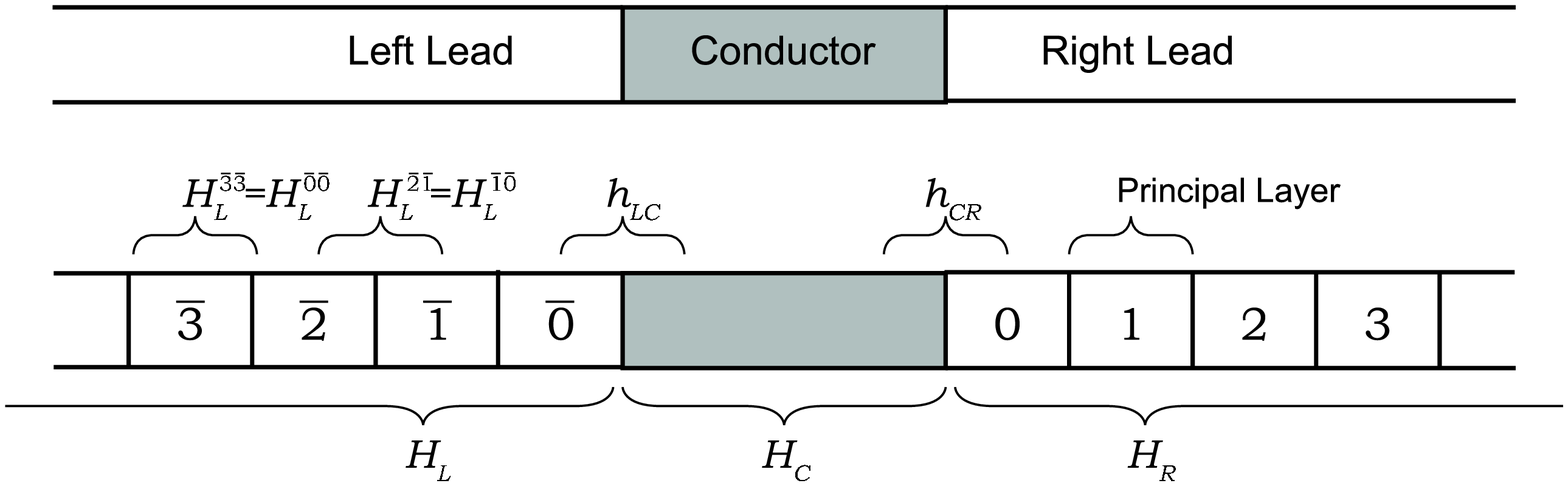}
\caption{Top: Schematic illustration of the lead-conductor-lead
  geometry. Bottom: An illustration of how the leads are split into
  principal layers with Hamiltonian sub-matrices labelled according to
  Eq.~(\ref{eq:split_hams}).}
\label{fig:l-c-l}
\end{figure}

Notwithstanding concerns regarding its ability to
describe charge transport in certain
situations~\cite{koentopp_2008}, DFT combined with
the Landauer formulation~\cite{landauer_1970} has become a standard
starting point for evaluating quantum conductance
(QC)~\cite{nardelli-prb01, taylor-prb01, brandbyge-prb02,
  wortmann-prb02, thygesen-prb03, calzolari-prb04, lee-prl05, 
  thygesen-cp05, polizzi-jcp05, havu-jcp06}. 
Calculations typically adopt a `lead-conductor-lead'
geometry (Fig.~\ref{fig:l-c-l} (top)) whereby
the conductor is sandwiched between two contacts, or
leads, whose semi-infinite nature is accounted for by means of surface
Green's functions and self-energies~\cite{datta_1995} 
obtained from standard DFT calculations. 

Despite the success of this approach, realistic nanoscale
systems, which typically contain arbitrary distributions of impurities,
functionalizations and modulations of structure and composition are 
challenging to describe accurately due to the asymptotic cubic scaling of
conventional DFT calculations with respect to system size. 

In this Article, following \citet{lee-prl05} and
\citet{cantele-nl09}, we use a method based on the
transferability of maximally-localised Wannier functions
(MLWFs)~\cite{marzari_1997,souza_2001} in order to overcome the cubic-scaling
bottleneck. 
The novelty of our work lies in the
development of robust algorithms for the complete automation of the
often painstaking manipulations required for preparing a
Hamiltonian matrix in the MLWF basis. As a result, high-throughput
computations of QC requiring little user intervention become feasible
for disordered nanoscale systems. 
Two further important features of our method are (i) that
the MLWF basis is optimally compact, ensuring highly efficient
determination of QC and density of states (DoS), and (ii)
that the nearsightedness of the electronic interactions can be
exploited in the MLWF basis by piecing together, without loss of
accuracy, Hamiltonians from DFT calculations on small fragments to
form model Hamiltonians of complex nanostructures consisting of tens
of thousands of atoms or more. 

%
 
The remainder of this paper is structured as follows:
Sec.~\ref{sec:methods} describes briefly the underlying theory of
Landauer transport and MLWFs, the real-space basis in which the
transport calculations are performed; Sec.~\ref{sec:SSG} describes the
details of the implementation of our automated method within the  
Wannier90 code~\cite{mostofi_2008}; in Sec.~\ref{sec:results}
we present the results of our approach on a number of systems; finally,
Sec.~\ref{sec:conclusion} is reserved for our concluding remarks.

\section{Theoretical Background}
\label{sec:methods}

\subsection{Landauer Transport}
\label{subs:landauer}

Within the Landauer formalism, it is assumed that there are no
dissipative scattering events on the length scale of the conductor
region, such that transmission is coherent, or ballistic. For a single
conducting channel at each energy $E$, Landauer
showed~\cite{landauer_1970} that the zero-bias, zero-temperature
conductance $G(E)$ is given by 
\begin{equation}
G(E)=\frac{2e^{2}}{h}T(E),
\label{eq:landauer_1}
\end{equation}
where $T(E)$ is the probability of transmission through the conducting
channel. In this framework, $G(E)$ is called the \emph{quantum
  conductance} (QC). Extending this formalism to multiple
channels~\cite{fisher_1981,anderson-prl80,meir-prl92} one may write
\begin{equation}
G(E)=\frac{2e^2}{h}\mathrm{Tr}(\Gamma_L G_{C}^{r} \Gamma_R G_{C}^{a}) ,
\label{eq:landauer_2}
\end{equation}
where $G_{C}^{\{r,a\}}$ are the retarded (r) and advanced (a) Green's
functions associated with the conductor, and $\Gamma_{\{L,R\}}$ are
functions that describe the coupling of the conductor to the left (L) and
right (R) leads.

The standard approach used to determine the QC of nanostructures 
that has emerged in recent years employs a localised basis set so that
the Hamiltonian $H$ of a system in the lead-conductor-lead geometry
(Fig.~\ref{fig:l-c-l} (top)) may be partitioned unambiguously. A
 \emph{principal layer}~\cite{lee_1981,lee2_1981} (PL) is
introduced, which is long enough so that  
$\exv{\zeta_i^n}{\hat{H}}{\zeta_j^{m}}\simeq0$
if $|m-n|\ge 2$, where $\zeta_i^n$ is the $i^{\mathrm{th}}$ basis
function in the $n^{\mathrm{th}}$ PL and $\hat{H}$ is the Hamiltonian
operator for the entire system. By imposing the equality on the Hamiltonian
elements, a truncation error is introduced, which is controlled systematically by
increasing the size of the PL (as will be shown in Sec.~\ref{sec:results}). 
The Hamiltonian matrix, with
reference to the bottom panel of Fig.~\ref{fig:l-c-l}, then takes
tri-block diagonal form, 
\begin{equation}
H=\left( \begin{array}{ccccccc}
\ddots & \vdots & \vdots & \vdots & \vdots & \vdots & \iddots \\
\cdots & H_L^{\bar{0}\bar{0}} & H_L^{\bar{1}\bar{0}} & 0 & 0 & 0 &
\cdots \\
\cdots & H_L^{\bar{1}\bar{0}\dag}	& H_L^{\bar{0}\bar{0}}& h_{LC}
& 0 & 0 & \cdots \\
\cdots & 0 & h_{LC}^\dag & H_C & h_{CR} & 0 & \cdots \\
\cdots & 0 & 0 & h_{CR}^{\dag}	& H_R^{00} & H_R^{01} & \cdots \\
\cdots & 0 & 0 & 0 & H_R^{01\dag} & H_R^{00} & \cdots \\
\iddots & \vdots & \vdots & \vdots & \vdots &\vdots & \ddots \\
\end{array}\right),
\label{eq:split_hams}
\end{equation}
where interactions between the first PL of the left or right lead and the
conductor are $h_{LC}$ and $h_{CR}$, respectively;
$H_{L}^{\bar{0}\bar{0}}$ and $H_{R}^{00}$ are matrices formed by
orbitals in the same PL in the semi-infinite left and right leads,
respectively, and $H_{L}^{\bar{1}\bar{0}}$ and $H_{R}^{01}$ are
matrices formed by orbitals in adjacent PLs in the left and right
leads, respectively. As shown in Eq.~(\ref{eq:split_hams}), these
latter four matrices are periodically repeated to form $H_L$ and $H_R$
(defined in the bottom panel of Fig.~\ref{fig:l-c-l}).

\subsection{Green's Function Formalism}
\label{subs:GF_formalism}

Knowledge of the seven finite Hamiltonian sub-matrices
$H_L^{\bar{0}\bar{0}}$, $H_L^{\bar{1}\bar{0}}$, $h_{LC}$ $H_C$,
$h_{CR}$, $H_R^{00}$ and $H_R^{01}$ is sufficient to describe the
open system of Fig.~\ref{fig:l-c-l} and to calculate the QC from
Eq.~(\ref{eq:landauer_2}).
Following \citet{nardelli_1999}, in order to determine
$G_{C}^{\{r,a\}}$ and $\Gamma_{\{L,R\}}$, we first consider the Green's
function $G$ of the whole system,
\begin{equation}
(\epsilon-H)G=\mathbb{I} ,
\label{eq:gf_whole_sys}
\end{equation}
where $\epsilon=E+i\eta$ ($\eta\rightarrow 0$) for $G^r$. Since
$G^a=(G^r)^\dag$, in the following we focus on $G^r$ only and suppress
the superscript. 
From Eq.~(\ref{eq:gf_whole_sys}) it can be shown that~\cite{datta_1995}
\begin{equation}
G_C=(\epsilon-H_C-\Sigma_L-\Sigma_R)^{-1},
\label{eq:G_C}
\end{equation}
where $\Sigma_L=h_{LC}^\dag G^{00}_Lh_{LC}$ and $\Sigma_R=h_{CR}^\dag
G^{00}_Rh_{CR}$ represent self-energy terms due to the coupling of the
conductor to the leads.
$G^{00}_{\{L,R\}}$ are known as surface Green's functions and can be
computed efficiently via the iterative procedure of
\citet{sancho_1984}.
$G_C$ is related to the local density of states (DoS) $\mathcal{N}_C$
of the conductor by~\cite{datta_1995} 
\begin{equation}
\mathcal{N}_C(E)=-\frac{1}{\pi}\mathrm{Im}(\mathrm{Tr}[G_C(E)]).
\label{eq:gf_dos}
\end{equation}
Finally, the coupling functions $\Gamma_{\{L,R\}}$ are given
by~\cite{datta_1995}
\begin{equation}
\Gamma_{\{L,R\}}=i[\Sigma^r_{\{L,R\}}-\Sigma^a_{\{L,R\}}].
\label{eq:gf_coupling_functions}
\end{equation}

In the special case that the lead and conductor are identical and the
entire lead-conductor-lead system is translationally invariant, the
following simplifications can be made: $H_L^{\bar{0}\bar{0}} =
H_R^{00} = H_C$, and $H_L^{\bar{1}\bar{0}} = h_{LC} = h_{CR} =
H_R^{01}$. 
Such systems are hereafter referred to as \emph{bulk}, or
\emph{pristine}, systems and transport calculations thereupon are
referred to as bulk, or pristine, transport calculations. In our
results, we will compare the QC of disordered conductors with the
corresponding bulk, or pristine, QC.

\subsection{MLWF Basis}
\label{subs:MLWF}

In a periodic crystal, within the independent particle approximation,
electrons are described by bands, or Bloch states
$\ket{\psi_{n\k}}$ with band index $n$ and crystal momentum
$\k$. An entirely equivalent representation may be 
constructed in terms of Wannier functions $\ket{w_{n\R}}$,
obtained by Fourier transforming $\ket{\psi_{n\k}}$ 
in the pair of conjugate variables $\k$ and $\R$, where $\R$ labels
the lattice vector of the real-space cell in which the Wannier
function is centered. 
Unlike Bloch functions, however, Wannier functions may be constructed
 that  are localized in real space, exhibiting exponential
decay in systems with an electronic band-gap~\cite{brouder_2007}.
Even in a metal, exponential localisation can be achieved if an appropriate
combination of filled and empty states is used\cite{lee-prl05}.

For an isolated band, observables are invariant under a gauge
transformation of the form $\psi_{n\k} \rightarrow
e^{i \phi_{n\k}}\psi_{n\k}$. Different choices of the phase
$\phi_{n\k}$, however, will result in different Wannier functions, and
can therefore be used as a means of making the resulting Wannier
function as localised as possible. 
\citet{marzari_1997} showed that for a composite, yet isolated group
of bands (such as those found in the valence manifold of an insulator
of semiconductor), one may define a set of generalised Wannier
functions
\begin{equation}
\ket{w_{n\R}}=\frac{V}{(2\pi)^3}\int_{BZ}
\sum_{m}^{N}U_{mn}^{(\k)} \ket{\psi_{m\k}}
e^{-i\k\cdot\R}d\k,
\label{eq:wf_def_isolated_bands}
\end{equation}
where $U_{mn}^{(\k)}$ is a unitary matrix
that may be chosen such that the Wannier functions are
\emph{maximally-localised} i.e. that the sum of their
quadratic spreads 
\begin{equation}
\Omega=\sum_{n}^{N}[\langle r^2\rangle_n-\langle \mathbf{r} \rangle_n^2],
\label{eq:wf_spread}
\end{equation}
where $\langle r^2\rangle_n=\langle w_{n\0}|r^2|w_{n\0}\rangle$
and $\langle\mathbf{r}\rangle_n=\langle
w_{n\mathbf{0}}|\mathbf{r}_n|w_{n\mathbf{0}}\rangle$, takes
the smallest value possible.

When a set of bands is not isolated from the rest of the band
structure by a gap across the Brillouin zone, the bands are said to be
connected or \emph{entangled}. This is the case in metals and in
conduction manifolds of semiconductors and insulators. In such cases,
within a given energy window, the number of bands at each point in
k-space varies and the disentanglement procedure of \citet{souza_2001}
is used in order to extract, or disentangle, an optimally-connected
subspace of a given, constant dimension at each $\k$. Once this
optimal subspace has been obtained, the usual localization procedure
of \citet{marzari_1997} may be applied in order to determine the MLWFs.
Once obtained, these provide a real-space and often intuitive picture
of bonding in materials, to the point that they are now used widely as a 
post-processing tool in electronic structure calculations~\cite{mostofi_2008}. 

There are a number of advantages to using MLWFs. First, they span a much smaller
subspace compared to, say, the plane-wave basis in which the original
ground-state electronic structure calculation is performed. 
The space of MLWFs is arguably the most compact, minimal manifold
possible (1 MLWF per every band that needs to be described), while
still preserving in full the accuracy of the electronic structure
calculation. As a
result, matrices in an MLWF basis can be orders of magnitude smaller
in each dimension than in the original basis, while still reproducing exactly
the properties of the ground-state, thus enabling very
efficient and accurate computation of ground-state properties, such as
interpolated band structures~\cite{yates_2007}. For example, the band structure
of the valence manifold for silicon is equivalently described by $\sim$3000 plane-waves
per atom or 2 MLWFs per atom. Second, since they
are localised in real-space, MLWFs may be used to represent the
Hamiltonian of a system in sparse matrix form. Finally, this sparsity
may be exploited in order to build model Hamiltonians of large,
structurally complex systems from Hamiltonians of smaller fragments.

\section{The Single Supercell Geometry}
\label{sec:SSG}

\begin{figure}
\centering
\includegraphics[height=4cm]{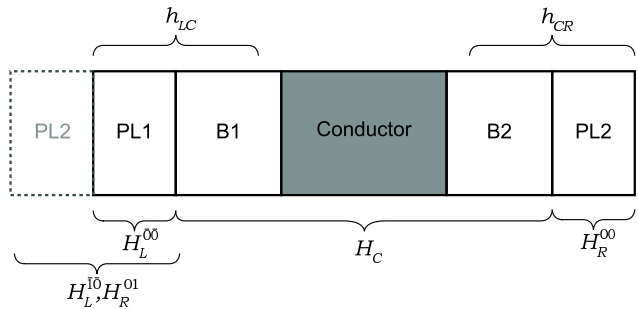}
\caption{Schematic illustration of the SSG (the single DFT
  supercell required for automated QC calculations). The conductor
  under investigation is flanked on each side by principal layers
  PL1, PL2 of the leads and a buffer region B1, B2. The buffer is a
  length of lead at least as large as a principal layer whose function
  is to ensure the disorder of the conductor has no significant effect
  on the periodicity of the lead Hamiltonian in PL1 and PL2. Also shown is the 
  periodic image of PL2 and the regions where each Hamiltonian 
  sub-matrix is derived from when expressed in the MLWF basis.} 
\label{fig:SSG}
\end{figure}

The translational symmetry present in crystals is exploited in
electronic structure calculations by using supercells and periodic
boundary conditions (PBC). A natural basis set to use for such
calculations is that of plane-waves and their benefits for DFT
calculations are well-understood~\cite{payne_1992}.

The lead-conductor-lead geometry of Fig.~\ref{fig:l-c-l}, however, is
inherently non-periodic. Therefore, as outlined in
Sec.~\ref{sec:methods}, if we are to use PBC for our Landauer conductance
calculations, a transformation
to a localised basis set, such as MLWFs, becomes essential.

Furthermore, the vast range of structural combinations
that one could investigate means that the change of basis must be
coupled to a robust and user-friendly algorithm that automatically
prepares the Hamiltonian obtained from a calculation on a periodic
system for use in the transport calculation so that high-throughput
calculations are possible. The novelty of our work lies in the
automation of these non-trivial manipulations of Hamiltonian matrices
and in streamlining the calculations such that
a calculation on only a single supercell is required; we call this
the \emph{Single Supercell Geometry} (SSG).

The SSG is shown in Fig.~\ref{fig:SSG}, whereby a central conductor is 
sandwiched between a length of lead on the left and right. The
conductor is the disordered region under investigation and the leads
are the contacts whose bulk is periodically repeated \emph{ad infinitum} in
the open (lead-conductor-lead) system. We split each lead into two parts: 
the outermost regions must be a PL of lead 
(PL1 and PL2) and  the inner regions a buffer (B1 and B2) such that any
disorder within the electronic structure associated with the conductor  
is localised within the region marked $H_{\rm C}$. 
In this respect it is important to converge results with respect to
the size of the buffer regions; we also impose that B1 and B2 must be at 
least one PL of lead in length.

By transforming to a MLWF basis, our algorithm uses the SSG to
identify the Hamiltonian sub-matrices required for the transport calculation.
Fig.~\ref{fig:SSG} depicts these regions. For
the sake of clarity, it is worth highlighting first that $H_C$ is in
fact built from the contribution of MLWFs within the conductor and the
buffers, and second that the interaction between two adjacent PLs of
lead, $H_L^{\bar{1}\bar{0}}$ and $H_R^{01}$, are built from
Hamiltonian matrix elements between MLWFs in PL1 and the periodic
image of PL2. For this reason, we demand that the left and right leads
of the SSG be identical in nature.

The Hamiltonian sub-matrices attained from partitioning the 
total Hamiltonian require a number of operations performed on them
before they can be input into transport calculations. 
First, we need to re-order the MLWFs in real-space so that every unit 
cell in PL1, PL2, B1 and B2 has a consistent sequence of MLWFs. 
This is because the Hamiltonian corresponding to the semi-infinite leads is
constructed from sub-matrices extracted from the SSG Hamiltonian in 
the MLWF basis. The connection matrix
$H_L^{\bar{1}\bar{0}}$ is constructed
from the Hamiltonian matrix elements between MLWFs in PL1 and the
periodic image of PL2, whereas $H_L^{\bar{0}\bar{0}}$ is
constructed from PL1 only. These two matrices are then duplicated along the
block off-diagonal and block diagonal, respectively, of the
Hamiltonian of Eq.~(\ref{eq:split_hams}). In doing so, the implicit assumption
is that the sequence of MLWFs in the rows and columns of the Hamiltonian
sub-blocks are the same, which in general is not true.
To overcome this problem we use the positions of the MLWF centres in real-space
to order the elements of the Hamiltonian sub-matrices:
the MLWFs in each unit cell of lead are arranged first according to
their position along one
direction perpendicular to the transport direction, then in the other
perpendicular direction, and finally along the transport direction itself.
This ensures that the sub-matrices can be used consistently to build the
Hamiltonian of Eq.~(\ref{eq:split_hams}).

The shape of MLWFs are often chemically intuitive and display
atomic-like or bonding/anti-bonding orbitals. Thus if more than one
MLWF exists with precisely the same centre, as may happen with
$d$-like MLWFs on a transition metal site, a second level of
ordering based on the orbital character of the MLWF is performed,
employing a technique we have developed using spatially-dependent
integrals to deduce a unique signature for each MLWF (see
\ref{ap:signatures}).

In addition to the ordering of the MLWFs, a second consistency
criterion must also be imposed.
The issue stems from the fact that, although MLWFs
are always found to be real, they remain undetermined upto an overall
sign, or \emph{parity}. 
As with the issue of ordering the MLWFs, the procedure of 
building the Hamiltonian from sub-matrices implicitly  assumes
that the MLWFs in PL2 have the same parity pattern as those in PL1, 
which in general is not true.

To address this issue, we enforce a consistent parity pattern 
at the level of the unit cell of lead onto the ordered Hamiltonian
sub-matrices (PL1, B1, B2 and PL2). The parities of
the MLWFs in the leftmost unit cell of lead in the SSG supercell are used
as the template. By assessing the relative parity of MLWFs in
this unit cell compared to translationally equivalent MLWFs in the
other unit cells of the PLs and buffer regions, the pattern is
enforced throughout by multiplying by $\pm 1$, as appropriate. The
relative parities are determined by using the unique signature
associated with each MLWF. 

We outline three caveats that apply to the current implementation of 
the SSG method: (i) the Bloch states used as input for determining the
MLWF basis in the SSG are calculated at the $\Gamma$-point only; (ii) the 
lattice vectors of the SSG must form a orthorhombic set and the direction of
conduction must be in the $x$, $y$ or $z$ direction.
(iii) the system under investigation must be quasi-one-dimensional (although
the extension to bulk leads would be relatively simple to implement.)

\begin{figure}
\centering
\includegraphics[width=8cm]{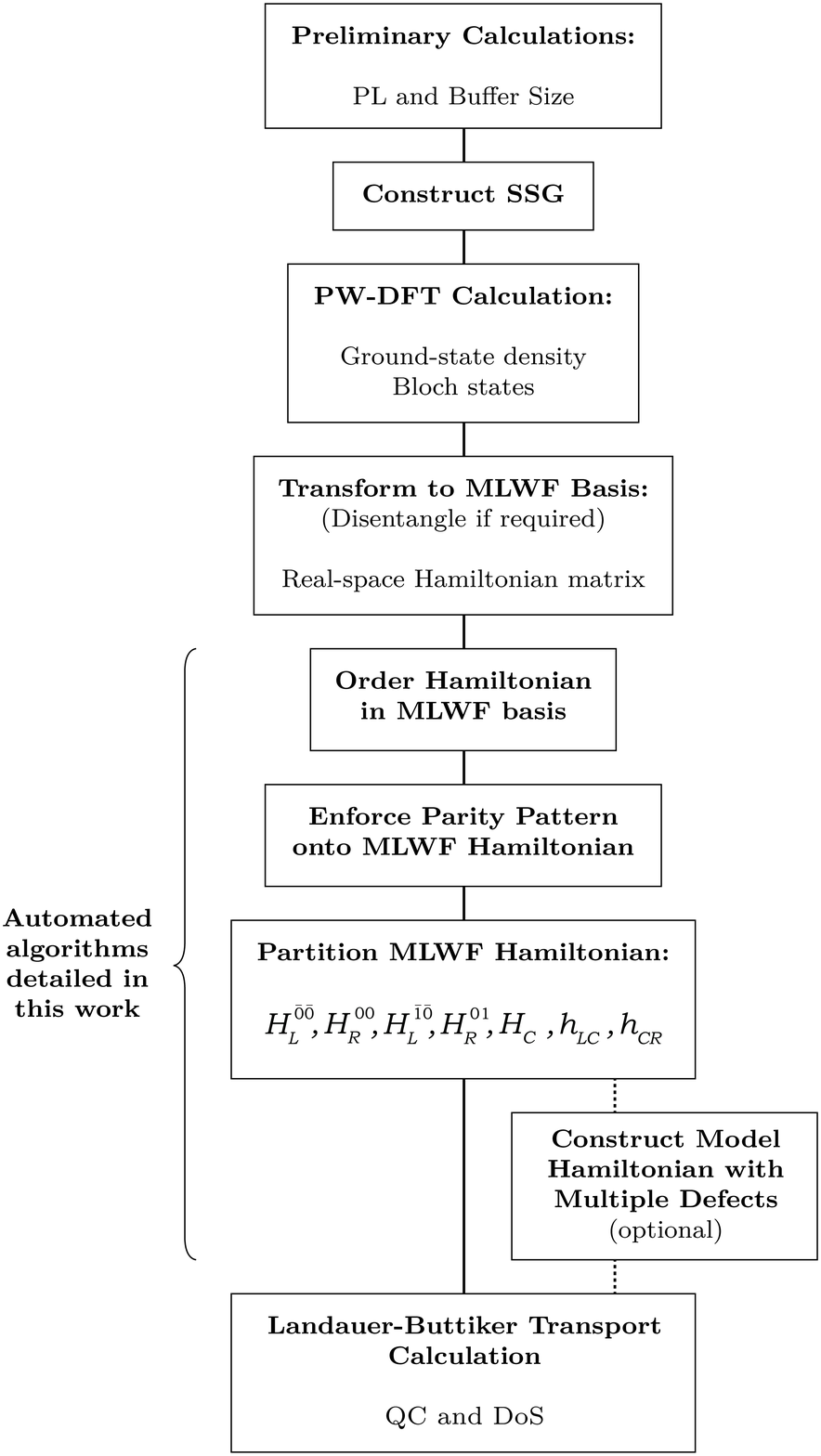}
\caption{A flow diagram depicting the key steps in our calculation procedure.}
\label{fig:flow_chart}
\end{figure}

\subsection{Calculation Procedure}
\label{sub:calc_procedure}

We now outline our general method, this is shown schematically in 
Fig.~\ref{fig:flow_chart}. First we must determine the number
of unit cells that make up a PL. Consider a supercell of lead with
$2n+1$ unit cells along the conduction direction, whose Hamiltonian in
the MLWF basis is found from a $\Gamma$-point DFT calculation in
PBC.
The value of $n$ is chosen such that Hamiltonian matrix elements
between MLWFs in the central unit cell and the left-most unit cell are
less than a certain threshold, which is usually set to be around 10 meV. 
In practice, for computational efficiency, instead of a $\Gamma$-point
supercell calculation, we apply Bloch's
theorem to reduce the supercell to a single unit cell, and perform the 
DFT calculation on a regular grid of k-points in the conduction
direction. 
This calculation is also used to calculate the bulk, or pristine, QC
which is used for validation purposes (see Sec.~\ref{sec:results}).

\begin{figure}
\centering
\includegraphics[width=8cm]{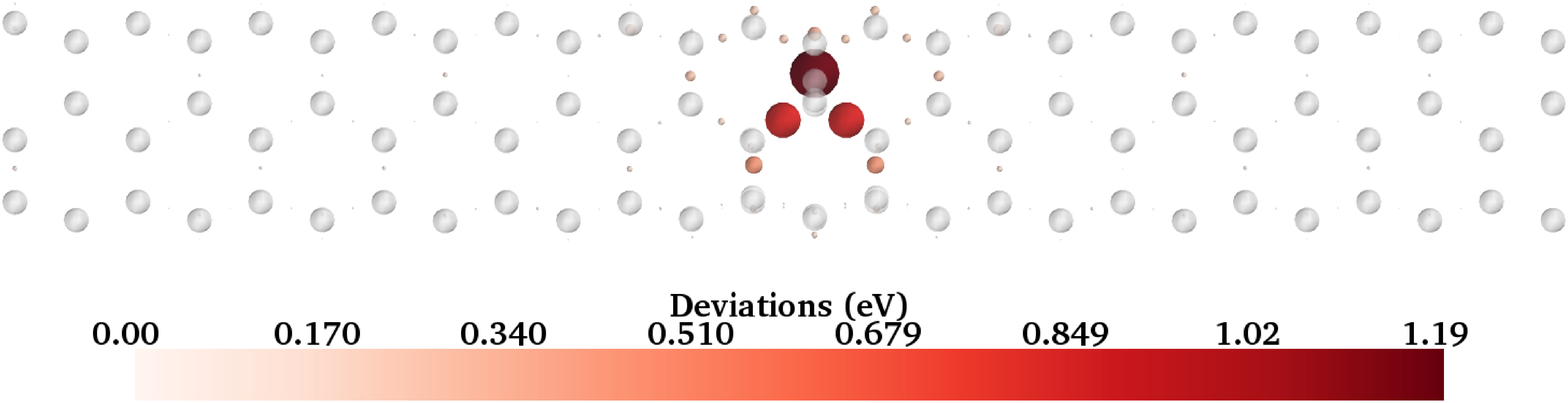}
\caption{Illustration of the inherent electronic nearsightedness in a (3,3) carbon
  nanotube functionalised with a single hydrogen atom. The white-gray spheres represent
  the atoms of the structure while the colored spheres represent the deviations from
  the ``bulk" values of the on-site Hamiltonian matrix elements for each MLWF. The size of
  colored spheres is another indication of the deviation of the matrix element from
  its ``bulk" value. The smaller the sphere, the smaller the deviation.}
\label{fig:defect_influence}
\end{figure}

Next, the extent of the buffer is determined by assessing the
convergence of electronic structure in PL1 and PL2 with respect to
its size. If the disorder present in the conductor region is
short-ranged, 
then often a single PL of lead in B1 and B2 is sufficient. This point
is illustrated in Fig.~\ref{fig:defect_influence}: we see that beyond
a few unit cells from a defect (hydrogen functionalization in a (3,3)
carbon nanotube), the on-site Hamiltonian matrix elements of the MLWFs
recover their bulk value.
Once the extent of the buffer and PLs have been determined, the SSG
supercell may be built and its Bloch eigenstates found from a
conventional DFT calculation. This calculation is usually performed in 
two steps. First, a self-consistent calculation at enough k-points to converge 
the charge density, followed by a non-self-consistent calculation at the $\Gamma$-point
only, using the self-consistent charge density as an input. 

Transformation to the MLWF basis,
Hamiltonian-matrix preparation, and transport calculations may then be
performed. The automated algorithms described above, which are
implemented in the Wannier90~\cite{mostofi_2008} code are designed so these steps
are performed sequentially, with little or no intermediary user input. A 
natural validation of the results may be performed whereby the
disordered conductor region of the SSG is replaced by a section of
pristine lead: identical results should be achieved with the bulk calculation.   

\subsection{Combination of multiple defects}
\label{subs:multiple_defect}

Moving from a Bloch to a Wannier representation is not only a means by
which to represent
electronic structure in a very compact manner. It also opens the
possibility to exploit the real-space nature of the basis to build
very large systems -- systems so large that a conventional DFT calculation
would be intractable.

The fact that electronic nearsightedness becomes explicitly manifest
in the MLWF basis, as highlighted in Fig.~\ref{fig:defect_influence},
allows them to be used to build the Hamiltonian matrix of a large
structure from the smaller Hamiltonian matrices of its constitutive
sub-systems.

\begin{figure}
\centering
\includegraphics[width=8.0cm]{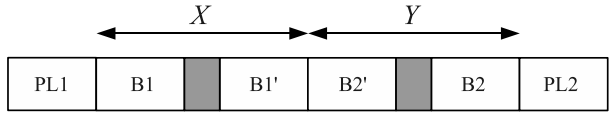}
\caption{Schematic for a SSG with a conductor containing two identical defects. 
We identify two regions in the conductor, $X$ and $Y$.}
\label{fig:2_defects}
\end{figure}

\begin{figure}
\centering
\includegraphics[width=7.0cm]{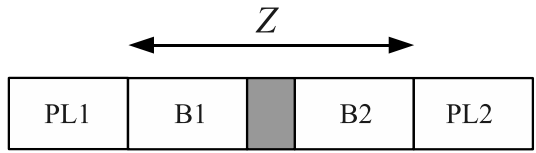}
\caption{Schematic for a SSG with a conductor $Z$ containing one defect.} 
\label{fig:1_defect}
\end{figure}

In order to illustrate the method, consider the schematic lead-conductor-lead 
system shown in Fig.~\ref{fig:2_defects} in which the conductor region
has two identical defects separated by a region of lead material in the form of a buffer 
(B1$^\prime$ and B2$^\prime$).
We could calculate the QC of this structure by making a SSG with
the whole conductor (regions $X$ and $Y$). However, we may exploit the nearsightedness
of the MLWF basis to find a more computationally efficient
approach. If the effect of the defects is localised (in the sense that the local
electronic structure and geometry at the junction between B1$^\prime$ and B2$^\prime$ 
is sufficiently similar to that seen in the leads),
then we may construct the Hamiltonian for the system with two defects
(Fig.~\ref{fig:2_defects}) from information gathered from one SSG calculation containing
just a single defect (Fig.~\ref{fig:1_defect}). Since this system is smaller, 
there is a clear advantage in terms of computational cost for the initial DFT calculation.

The Hamiltonian of the conductor in the two-defect system may be written as
\begin{equation}
H_C=\left( \begin{array}{cc}
H_X				 			& H_{XY}	\\
H_{XY}^{\dagger}	& H_Y 			\\
\end{array}\right)
\end{equation} 
where $X$, $Y$ and $XY$ represent blocks of Hamiltonian matrix elements
among MLWFs in region $X$, among MLWFs
in region $Y$, and between MLWFs in these two regions, respectively.

Given the geometry of the system, and the nearsightness of the electronic structure,
blocks $H_X$ and $H_Y$ should be quite close in terms of their matrix elements. 
Moreover, because of the constraint that a buffer is at least as large as a lead principal 
layer, we expect  the non-zero matrix elements of $H_{XY}$ to correspond closely
to the overlaps between the two adjacent principal layers.
This observation stems from the very definition of a principal layer. As a consequence, 
we can construct a close approximation to $H_C$
by using the matrices extracted from a SSG calculation of the structure shown in
Fig.~\ref{fig:1_defect}. In this approximation, blocks $H_X$ and $H_Y$ are replaced 
with $H_Z$, and $H_{XY}$ is replaced by the overlap matrix between two principal layers
of lead (namely $H_L^{\bar{1}\bar{0}}$, see Fig. \ref{fig:SSG}), i.e.,
\begin{equation}
H_C\simeq\left( \begin{array}{cc}
H_Z				 									& H_L^{\bar{1}\bar{0}}	\\
H_L^{\bar{1}\bar{0}\dagger}	& H_Z 									\\
\end{array}\right).
\end{equation}
An example of this approach is demonstrated in Sec.~\ref{sec:results} for a defected
silicon nanowire.

The approach described above is general and may be applied to any number of
isolated defects in the conductor region. In this way, Hamiltonians
for systems of almost arbitrary size may be constructed with
first-principles accuracy from one DFT calculation in a SSG with a
single defect. We note in passing the importance
that the MLWFs parities are consistent between different
regions of the system. As mentioned in Sec.~\ref{sec:SSG}, the parities need to 
be checked and made consistent to allow
seamless connections between Hamiltonian sub-matrices, a task
that is automatic in the present approach.

Furthermore, the Hamiltonian of a conductor with more than one
type of defect may be constructed by combining matrix elements from
separate SSG calculations. In this latter case, care must be taken in order 
to align the Fermi energies of the two (or more) distinct calculations. 
This is the consequence of the lack of an absolute reference for the electrostatics
in PBCs, which can lead to Fermi energies that are shifted by a constant.

Additionally, building a large-scale structure with tens of defects and tens of thousands of atoms
can be a painstaking task. In order to simplify this process, we have
designed a utility package to the Wannier90 code that helps the user
create these large scale structures. From a single Wannier90 
calculation in the SSG geometry, both randomised and custom-made structures can be built
with ease. An illustration of the use of that functionality is given in the fourth example of 
Sec.~\ref{sec:results}.

\section{Applications}
\label{sec:results}

We present now a number of examples using the method described in
Sec.~\ref{sec:methods}. The aim is to illustrate its robustness 
in a range of applications: beginning with a defected atomic chain, and
building complexity via a heterostructured nanowire and a
spin-polarised graphene nanoribbon. Finally, we provide 
an example to validate the use of SSG Hamiltonian fragments
in the construction of model Hamiltonians for larger systems. All 
DFT calculations are performed with the Quantum-ESPRESSO
package\cite{giannozzi_2009} and with (unless otherwise stated)
norm-conserving pseudopotentials.\cite{troullier_1991}

\subsection{Atomic Al Chain}

First, we consider the QC of an Al chain with a single Na atom substitutional defect. 
The construction of the SSG is performed with care:
a suitable PL length must first be decided upon by
assessing the rate of the decay of the matrix elements of the Hamiltonian
between MLWFs. Additionally, the defect
is expected to have a large effect on the electronic structure,
thus the buffer size must also be carefully chosen.  

To assess the length of a PL we use the method outlined in the
Sec.~\ref{sub:calc_procedure}, whereby the Hamiltonian in the MLWF
basis of a single unit cell of lead is determined at many
k-points. 
We perform the DFT calculation on a single unit cell (consisting of
one Al atom), with a regular grid of 32 k-points along the extended
direction. The total energy is converged to $10^{-11}$~eV using a
500~eV energy cut-off, exchange and correlation are described by the 
PBE functional\cite{perdew_1996}. The unit cell is 2.47~\AA\ long in the
conduction direction, with 10~\AA\ separating periodic images in the 
transverse directions. We proceed to the determination of the 
MLWF basis by disentangling three Wannier functions 
from 30 bands.

\begin{figure}
\centering
\includegraphics[height=5.5cm]{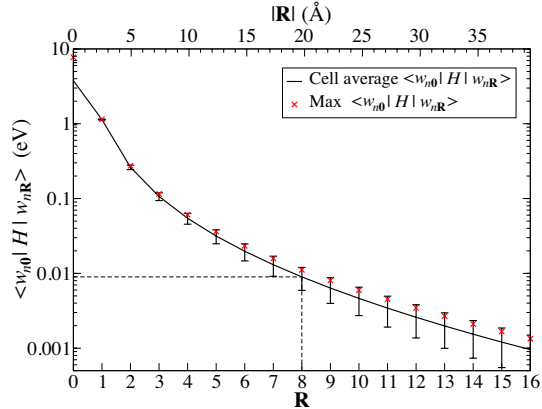}
\caption{Decay of the Hamiltonian elements 
  $\langle w_{n\0} \vert \hat{H} \vert w_{n\R} \rangle$
  between increasingly distant Al unit cells in units of the primitive
  lattice vector.
  Cell averaged elements are shown in black (error bars
  show a standard deviation); the largest Hamiltonian values between
  unit cells increasingly distant from $\R=0$ are shown by red
  crosses. The dashed line highlights the chosen PL size (see text).
} 
\label{fig:al_decay}
\end{figure}

Fig.~\ref{fig:al_decay} shows the decay of the interaction 
between MLWFs by averaging the on-site Hamiltonian elements $\langle
w_{n\0} \vert \hat{H} \vert w_{n\R} \rangle$ between
equivalent MLWFs in different unit cells labelled by the primitive
lattice vector $\R$ 
(black solid line). The maximum matrix element
(red crosses) gives the maximum error incurred due to truncation of
the interaction if the PL were to be cut at that unit cell. 
In this case, the PL is chosen to be eight unit cells long, such that
the maximum truncation error is approximately 11~meV and the average
error is 9~meV.

\begin{figure}
\centering
\includegraphics[height=5.5cm]{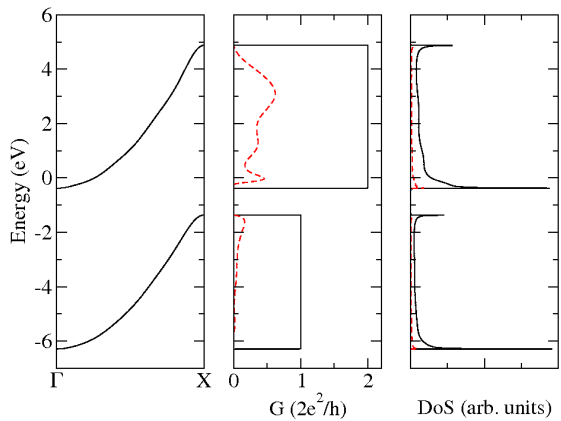}
\caption{Left: Band structure of the single cell, single Al atom bulk
  transport calculation. Centre: QC comparison of bulk (black, solid)
  and defected SSG (red, dashed) systems. Right: DoS comparison of bulk
  and defected SSG.} 
\label{fig:al_3_graphs}
\end{figure}

A buffer size of one PL plus three unit cells is chosen for the SSG
such that on relaxation the RMS difference in the position of the MLWF
centres from  
their ideal bulk position in the rightmost unit cell of PL1 is less
than $5\times10^{-3}$~\AA. The SSG therefore consists of a total of 39
atoms. Performing the transport calculation provides the QC and DoS
shown in Fig.~\ref{fig:al_3_graphs} (red, dashed; centre and right
panels, respectively). 
For comparison, the bulk band structure, QC and DoS (black; left, centre and
right panels, respectively) are also shown. In the bulk case there are clear
contributions from the $s$ band and two degenerate $p$ bands to the QC:
these are both significantly reduced in the defected case, with
conductance close to zero at lower energies. This may be interpreted
as the hybridization of the $s$ orbital associated with the Na to
adjacent Al $p$ orbitals. 

\subsection{Si/Ge Nanowire Heterostructures}
\label{subs:het_nw}

\begin{figure}[h]
\centering
\includegraphics[height=3.8cm]{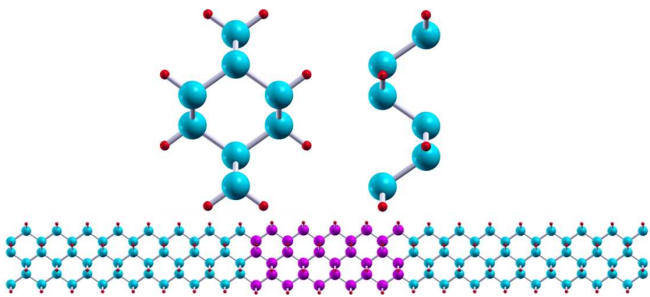}
\caption{Top: Front and side view of single Si cell used for PL
  determination and bulk transport calculations. Bottom: Supercell of
  the Si/Ge nanowire for use in the SSG method. Red,
  cyan and magenta represent H, Si and Ge atoms respectively.}
\label{fig:sinw_cells}
\end{figure}

\begin{figure}
\centering
\includegraphics[height=5.5cm]{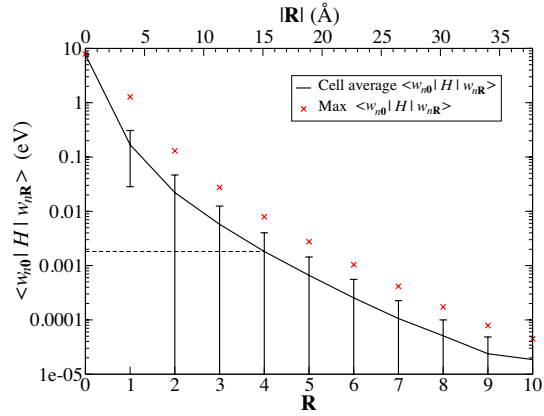}
\caption{Decay of the Hamiltonian elements 
$\langle w_{n\0} \vert \hat{H} \vert w_{n\R} \rangle$
 between increasingly distant Si nanowire unit cells.
  The notation used is equivalent to that of
  Fig. \ref{fig:al_decay}, where average on-site elements are shown in
  black (error bars show a standard deviation) and the largest
  Hamiltonian value between unit cells increasingly distant from unit
  cell 0 are shown by red crosses. The dashed
  line highlights the chosen PL size (see text).
}
\label{fig:sinw_decay}
\end{figure}

We now increase the complexity of the SSG system by considering 
a thin (0.39~nm radius) Si nanowire in the $\langle 110 \rangle$
direction with a Ge heterostructure inserted
as a defect. The DFT calculations detailed in this example (and those
on the nanowires of in Sec. \ref{subs:double_defect})
are performed within the
LDA and with an energy cutoff of 400~eV. 
We begin with a single cell for the lead (8 Si atoms, 8 H atoms) (see
Fig.~\ref{fig:sinw_cells} (top)) and perform a DFT calculation with
20 k-points, allowing atomic positions and lattice parameter in the
conduction direction to relax. Forces are converged to 5~meV/\AA.
Once the ground-state is found, we
transform to the MLWF basis and assess the PL length in the usual
manner. Fig.~\ref{fig:sinw_decay} displays the decay of the
Hamiltonian matrix elements as a function of increasingly distance,
using the same notation as Fig.~\ref{fig:al_decay}. 
With four unit cells in a PL the average truncation error is below
2~meV.

\begin{figure}
\centering
\includegraphics[height=5.3cm]{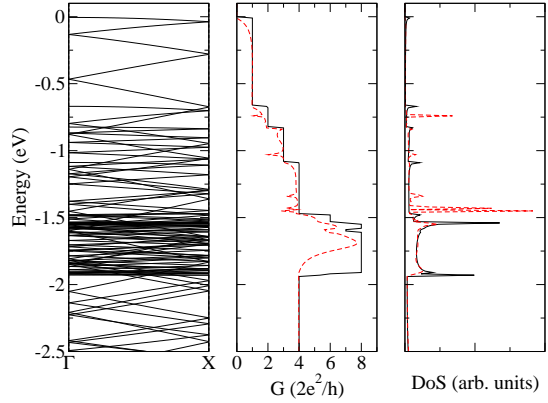}
\caption{QC (centre panel) and DoS (right panel) for pristine Si
  nanowire (solid, black lines)
 and axially heterostructured Si/Ge nanowire (dashed, red
 lines). The bandstructure of the pristine silicon nanowire is also
 shown (left panel).} 
\label{fig:sinw_3_graphs}
\end{figure}

A SSG is built by repeating the single unit cell of Si and inserting 
5 copies of a similarly relaxed Ge unit cell (see Fig.~\ref{fig:sinw_cells}, 
bottom panel). Without further relaxations of the geometry, it was found 
that a single PL was sufficient to converge the
electronic structure in PL1 and PL2 to that of  a bulk lead. Hence our
SSG consisted of 16 Si unit cells with five Ge unit cells sandwiched at
their centre (Fig.~\ref{fig:sinw_cells}, bottom panel). Using our
automated routines, this 336 atom unit cell provides the valence QC and DoS
shown in Fig.~\ref{fig:sinw_3_graphs} (red, dashed lines; centre and
right panels, respectively). 
The bandstructure, QC and DoS of the pristine silicon nanowire are
also shown (black solid lines; left, centre and right panels,
respectively). The drop-off in conductance just below the Fermi level
is due to localization of the highest occupied molecular orbital
within the Ge quantum well.

\subsection{Spin-polarised graphene nanoribbon}

In this third example, we look at a spin-polarised graphene nanoribbon functionalised
with a single hydrogen atom. As with the previous examples, we start with
a calculation on a single unit cell. Our system is a zigzag nanoribbon
of length 2.46~\AA\ with a width of 9.27~\AA. A regular grid of 20 k-points 
is used in the conduction direction and the supercell is built such that a vacuum
region of 10~\AA\ lies between periodic images. A cutoff of 400~eV for the
kinetic energy and 4500~eV for the charge density is used together with
a PBE exchange and correlation functional and ultrasoft 
pseudopotentials\cite{vanderbilt_1990}. Both the atomic positions 
and the unit cell length were fully relaxed;
individual forces are less than 18~meV/\AA. 

We must specify a starting non-zero magnetization
such that the self-consistent loop converges to a magnetic 
state (in this case we restrict ourselves to a ferromagnetic state, even though the ground
state is anti-ferromagnetic~\cite{louie_nature2006}). Next, a non-self consistent calculation 
is used to compute the band energies for both ``up" and ``down" spins. 
It is important at this stage to compute a sufficient number of bands
to capture the entire $\pi$ manifold,
otherwise the disentanglement of the conduction manifold would be
meaningless. In our calculations, we used 30 bands and we kept all the
bands up to $-0.5$~eV in the frozen window  for the disentanglement
procedure.

\begin{figure}[t!]
\centering
\includegraphics[height=5.3cm]{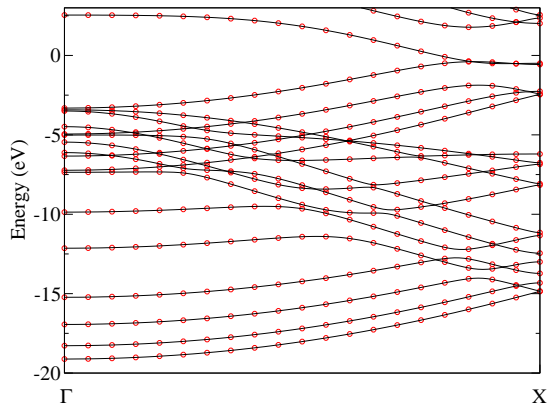}
\caption{Interpolated band structure (up spins) for a pristine zigzag
  graphene nanoribbon (black solid lines), 
and ``exact'' band structure in the complete plane-wave basis set 
given by the electronic structure code
(red dots). The Fermi energy is set to 0 eV.}
\label{fig:zgnr_bands}
\end{figure}

The quality of the disentanglement may be assessed by comparing the interpolated band structure 
provided by Wannier90 to the full band 
structure given by the electronic structure code (for this ferromagnetic state, both spin types
have an almost identical band structure except around the Fermi
level). As can be seen from Fig.~\ref{fig:zgnr_bands}, the 
match between the Wannier interpolation (solid lines) and the ``true'' band structure (dots) is
excellent. We see that the interpolated band structure describes perfectly conduction states
upto about 2.5~eV above the Fermi energy.

Satisfied with the MLWF transformation, the PL size is assessed in the manner
described earlier. Choosing a PL size of four unit cells (with a
maximum truncation error of less than 46~meV) is sufficient for this example.

\begin{figure}
\centering
\includegraphics[width=8cm]{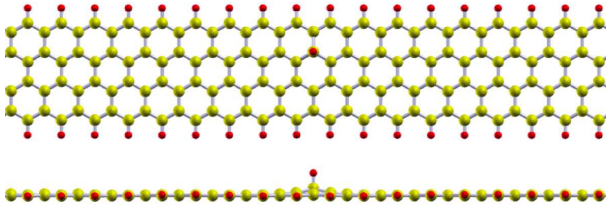}
\caption{Top and side view of the SSG structure used to compute the quantum conductance of both
 ``spin up" and ``spin down" channels. Carbon atoms are yellow and hydrogen is shown in red. The
 overall supercell consists of 18 unit cells of lead and an extra hydrogen, which totals 181 atoms.}
\label{fig:zgnr_supercell}
\end{figure}

The SSG consists of 4 unit cells of lead in both the PL1, PL2, B1 and B2
regions. Two unit cells form the conductor region (see Fig.~\ref{fig:zgnr_supercell}), upon
which a hydrogen atom is placed at an approximate C-H bond length above one of the carbon atoms. 
The whole system is fully relaxed both for atomic positions and supercell length in the 
direction of conduction. The force convergence criteria, cutoffs and energy convergence criteria 
are similar to the ones used in the single lead unit cell case above.

\begin{figure}
\centering
\resizebox{5cm}{!}{\includegraphics[6cm,1cm][20cm,22cm]{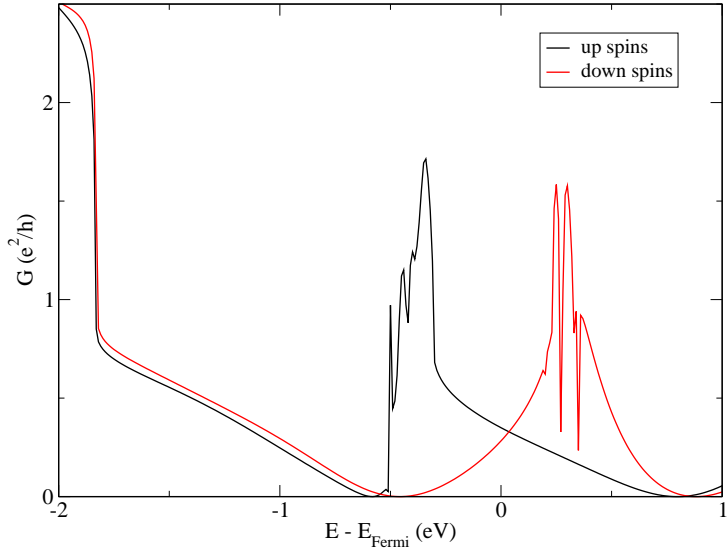}}
\caption{Spin-dependant QC close to the Fermi level. One can clearly see that depending
on the applied bias, a spin-polarised current can be induced in this system.}
\label{fig:spin_conductance}
\end{figure}

After relaxing the structure, the next step is to perform a spin-polarised DFT calculation followed by a
non-self consistent calculation to extract the ``up'' and ``down''
band energies at $\Gamma$.  
The QC is calculated in the usual manner for each spin channel separately with
Wannier90.
The result for the spin-dependent QC is shown in Fig.~\ref{fig:spin_conductance}.

We see on the graph that at the Fermi level the quantum conductance of
the system is slightly spin-polarised with a  majority of ``up''
spins. Applying a slightly negative bias, we see that the system 
can be close to 100$\%$ spin-polarised. The
opposite spin polarization can be achieved 
with a slightly positive bias.

\subsection{Doubly Defected Si Nanowire}
\label{subs:double_defect}

This final example demonstrates an extension to the SSG
method whereby the sub-Hamiltonians it creates are manipulated
and combined to construct model Hamiltonians of
larger systems (see Sec.~\ref{subs:multiple_defect}). All DFT 
calculations outlined here are performed within the LDA, at the $\Gamma$ point,
with the same energy cut off and tolerances described in Sec.~\ref{subs:het_nw}.

\begin{figure}[h]
\centering
\includegraphics[height=1.5cm]{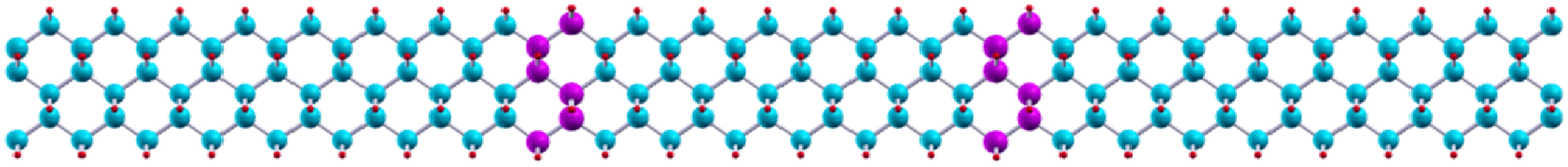}
\includegraphics[height=5.8cm]{double_defect_qc.eps}
\caption{Top: Si nanowire with two Ge heterostructure defects. 
The system is investigated by manipulating the Hamiltonian 
of a single defect (multiple defect method) and directly
using a SSG. Red, cyan and magenta atoms are H, Si and Ge 
respectively. Bottom: Comparison of QC for the two methods, 
showing excellent agreement.}
\label{fig:double_defect}
\end{figure}

The system we aim to describe is shown in Fig.~\ref{fig:double_defect}
(top): a Si nanowire with two single cells  
of Ge separated by a length of Si. This system may be thought of in
two ways: first, as a Si nanowire with a single defect containing 
the two cells of Ge and the separating Si cells; and second as 
doubly defected Si nanowire,
with each defect being a single cell of Ge. These two perspectives 
lead to two methods by which we can determine the QC. The first 
suggests a SSG calculation in which the conductor region 
contains the two Ge defects. The resulting QC is see in solid black 
in Fig.~\ref{fig:double_defect} (bottom). This is compared to 
the QC derived from a calculation in which the multiple defect method
described in Sec.~\ref{subs:multiple_defect} is used.  

For the doubly defected case, we perform a SSG calculation on a Si
nanowire with a single Ge cell defect and use the Hamiltonians
provided to build a Hamiltonian of the larger system in 
question (Fig.~\ref{fig:double_defect} (top)). The QC 
from this calculation is seen in dashed red in
Fig.~\ref{fig:double_defect} (bottom). The two calculations 
agree remarkably well, validating the premise of nearsightedness
discussed earlier. 
Since the computational expense of conventional DFT methods
scales as $O(N^{3})$, where $N$ is the number of 
atoms in the supercell, the multiple defect method 
represents a significant step forward to describe realistic system sizes with 
first-principles accuracy, and 
can be used to construct faithful model Hamiltonians 
 for systems that contain tens of thousands
of atoms\cite{shelley_2011,Li_Poilvert_2011}.

\section{Conclusions}
\label{sec:conclusion}
In this paper we have presented a user-friendly and automated
approach to calculate the quantum conductance and density of states in
quasi-one-dimensional systems. The method converts the Bloch
eigenstates of a single DFT calculation, within our single supercell
geometry, to a basis of MLWFs. In this basis we 
determine the electronic transport properties by automatically
extracting the Hamiltonian sub-matrices required for the transport
calculation. To illustrate the robustness, wide applicability and
efficiency of our method, we have presented calculations on an
atomic Al wire, a spin-polarised graphene nanoribbon, and axially heterostructured
Si/Ge nanowires. Furthermore, we have shown how the transport
properties of meso-scale conductors that are beyond the current
capabilities of conventional first-principles electronic structure
calculations can be calculated with first-principles accuracy by
exploiting the transferability of MLWFs as building blocks of large
model Hamiltonians.

\section*{Acknowledgements}
We would like to thank the Imperial College High Performance Computing
Service for continued use of their facilities and the developers of
the Quantum-ESPRESSO package~\cite{giannozzi_2009}. We are also
grateful to Davide Ceresoli and Young-Su Lee for their advice and help. Financial
support was provided by Research Councils UK, the UK Engineering and Physical Sciences
Research Council, The Institute for Soldier Nanotechnology at MIT and the Thomas
Young Centre for Theory and Simulation of Materials.

\appendix
\section{MLWF Signatures}
\label{ap:signatures}

Here we detail the set of spatially-dependent integrals that we use to determine 
a signature for each MLWF. These signatures are used for two purposes. 
First, they enable a sorting algorithm to distinguish between MLWFs of
different shapes with similar centers. Thus they may be ordered consistently over 
between unit cells -- a key requirement for our approach. Secondly, they are 
used to determine the relative parity of MLWFs so that a consistent parity-pattern
may also be enforced. 

We begin with the integral
\begin{equation}
I_n(\q)=\frac{1}{V}\int_{V} w_{n}(\mathbf{r}) \ee^{i\q\cdot(\mathbf{r}-\mathbf{r}_c)}\:d\mathbf{r} ,
\label{eq:sig_int_form}
\end{equation}
where $V$ is the volume of the cell, $\q$ is a vector in reciprocal 
space and $\mathbf{r}_{c}$ is the centre of Wannier function $w_{n}(\mathbf{r})$
(we assume sampling at $\Gamma$-point only). 
One may write $w_{n}(\mathbf{r})=\sum_m U_{mn} u_m(\mathbf{r})$, where 
$u_m(\mathbf{r})$ is the periodic part of the Bloch wavefunction
at band $m$. $U_{mn}$ is the unitary matrix found in 
Eq.~(\ref{eq:wf_def_isolated_bands}) that minimises the spread of the
Wannier functions. $u_m(\mathbf{r})$ can be written in terms of its
discrete Fourier transform $\tilde{u}_{m}(\G)$,
$u_{m}(\mathbf{r})=\sum_{\G} \tilde{u}_{m}(\G)e^{i\G\cdot\mathbf{r}}$.
Thus, the integral in Eq.~(\ref{eq:sig_int_form}) may be written as
\begin{equation}
I_n(\q)=e^{-i\q\cdot\mathbf{r}_{c}}\sum_{m} U_{mn} \tilde{u}_{m}^{\ast}(\q),
\label{eq:sig_fourier_form}
\end{equation}
where $\q$ is a $\G$-vector of the form 
$l\b_1+m\b_2+n\b_3$, where 
$\{l,m,n\} \in \mathbb{Z}$ and $\{\b_1,\b_2,\b_3\}$ are the
reciprocal lattice vectors. Equating real and imaginary parts of
Eq.~(\ref{eq:sig_int_form}) and Eq.~(\ref{eq:sig_fourier_form}), one
may write
\begin{align}
I_n^{\mathrm{Re}}(\q) &= \frac{1}{V} \int_{V}w_{n}(\mathbf{r}) 
\cos(\q\cdot(\mathbf{r}-\mathbf{r}_c))\:d\mathbf{r} \notag\\
&= \mathrm{Re}\left[
e^{-i\q\cdot\mathbf{r}_{c}} \sum_{m} U_{mn} \tilde{u}_{m}^{*}(\q) \right],
\label{eq:sig_real}
\end{align}
and
\begin{align}
I_n^{\mathrm{Im}}(\q) &= \frac{1}{V}\int_{V}w_{n}(\mathbf{r}) 
\sin(\q\cdot(\mathbf{r}-\mathbf{r}_c)) \:d\mathbf{r} \notag\\ 
&= \mathrm{Im} \left[
e^{-i\q\cdot\mathbf{r}_{c}} \sum_{m} U_{mn} \tilde{u}_{m}^{*}(\q)\right].
\label{eq:sig_imag}
\end{align}
Since most DFT codes compute $\tilde{u}_{m}(\G)$, obtaining 
any set of $I_n$ incurs negligible computational expense. 

The set of integrals that are used to determine 
a signature are given by
%
\begin{equation}
I_n=\frac{1}{V}\int_V w_n(\mathbf{r})
\sin^\alpha \left(\frac{2\pi}{L_x}(x-x_c)\right)
\sin^\beta \left(\frac{2\pi}{L_y}(y-y_c)\right)
\sin^\gamma\left(\frac{2\pi}{L_z}(z-z_c)\right)\:d\mathbf{r}
\end{equation}
%
where $\mathbf{r}_c=(x_c,y_c,z_c)$, $V=L_xL_yL_z$,  
$\alpha,\beta,\gamma \in \{0,1,2,3\}$ and  
$\alpha+\beta+\gamma \leq 3$. Each of the resulting 20 integrals may be written as
linear combinations of those outlined in
Eqs.~(\ref{eq:sig_real}) and (\ref{eq:sig_imag}).  
The signature of the MLWF is thus given by the 20-element unit vector
of these integrals. Dot products between two MLWFs' 
signatures reveal in a compact form their relative shape and parity.

\bibliographystyle{model1-num-names}



\end{document}